\documentclass[preprint2]{emulateapj}
\usepackage{graphicx}
\usepackage{times}
\usepackage{verbatim}
\usepackage{natbib}
\usepackage{amsmath} 
\usepackage{amsbsy}

\slugcomment{}

\shorttitle{Self-Perpetuating Spiral Arms}
\shortauthors{D'Onghia et al.}

\begin{document}

\title{Self-Perpetuating Spiral Arms in Disk Galaxies}

\author{Elena D'Onghia\altaffilmark{1,2,3}, Mark Vogelsberger\altaffilmark{1},
Lars Hernquist\altaffilmark{1}}

\altaffiltext{1}{Harvard-Smithsonian Center for Astrophysics, \\
60 Garden Str., Cambridge, MA 02138 USA}
\altaffiltext{2}{Keck Fellow; edonghia@cfa.harvard.edu}
\altaffiltext{3}{University of Wisconsin, \\
475 Charter St., Madison, WI 53706}

\begin{abstract}
The causes of spiral structure in galaxies remain uncertain. 
Leaving aside the grand
bisymmetric spirals with their own well-known complications, 
here we consider the possibility that
multi-armed spiral features originate from density inhomogeneities
orbiting within disks.  Using high-resolution N-body simulations, we
follow the motions of stars under the influence of gravity, and show
that mass concentrations with properties similar to those of giant
molecular clouds can induce the development of spiral arms through a
process termed swing amplification.  However, unlike in earlier work,
we demonstrate that the eventual response of the disk can be highly
non-linear, significantly modifying the formation and longevity of the
resulting patterns.  Contrary to expectations, ragged spiral
structures can thus survive at least in a statistical sense long after the
original perturbing influence has been removed.  

\end{abstract}


\section{Introduction}

Seventy percent of galaxies in the nearby Universe are characterized
by a disk with prominent spiral arms, but our understanding of the
origin of these patterns is incomplete, even after decades of
theoretical study \citep{Toomre77,Lia84,BT08,Sel11}.  Several ideas have been
proposed to explain the formation of spiral arms.  One model posits
that these features are large-scale density waves continuing to
propagate in a differentially rotating disk.  In particular, this
theory argues that the matter in the galaxy (stars and gas) can
maintain a density wave through gravitational interactions even in the
presence of shear.  This density wave remains at least
quasi-stationary in a frame of reference rotating around the center of
the galaxy at a fixed angular speed, identified with the pattern speed
of the spirals, and covers the entire disk
\citep{Lin64,BL96}.

An alternative theory proposes that spiral arms are stochastically
produced by local gravitational amplification in a differentially
rotating disk \citep{GLB65,JT66}.  The mechanism behind this process
is known as swing amplification and it can be seeded either 
by preexisting leading waves or else by 
the response
of a disk to the presence of a corotating overdensity, such as a giant
molecular cloud.    
This dynamical response takes the
form of wakelets in the surrounding medium, each 
amplified by its own self-gravity through the swinging of leading
features into trailing ones owing to the shear.

According to this second theory, spiral arms would fade away in one or two
galactic years if the driving perturbations were removed \citep{TK91},
in contrast to the quasi-steady nature of the arms in the model
proposed by Lin and Shu.  Thus, a continuous source of
perturbations would be required for these fluctuating spiral patterns
to be maintained throughout the lifetime of a galaxy.  Indeed, by
mimicking the effects of dissipative infall of gas, \citet{SelCarl84}
showed that the addition of fresh particles on circular orbits could
cause such spiral patterns to recur, and \cite{CF85} demonstrated that
almost any mechanism of dynamical cooling can maintain spiral
activity of a similar kind. 

However, the pioneering work on density waves in the 1960s was based
on analytic theory and invoked various assumptions in order for
solutions to be found.  For example, the swing amplification
analysis involved linear approximations to the equations of motion.
The theoretical emphasis since that time has been on identifying driving
mechanisms that can sustain the wave in spite of the damping which
would cause it to decay in this picture.  Indeed, whereas observations
show that spiral arms might be density waves, N-body experiments have
not yielded long-lived spiral structures, as predicted by
the stationary density wave theory.  Simulations of cool,
shearing disks always exhibit recurrent transient spiral activity and
this situation has not changed over the past several decades as
computational power has increased \citep{Sel00,Fujii11,Sel11}.

Some
work showed that spiral patterns fade away in numerical simulations of
stellar disks if the effects of gas dissipation are not included; the
reason is that the disk becomes less responsive as random motions rise
owing to particle scattering by the spiral activity and giant molecular clouds
\citep{SelCarl84,BL88}.  Moreover, the debate about the
longevity of the arms practically ceased two decades ago because the
available computational power did not permit definitive tests of some
of the predictions of the theories and also because observations at that
time were not sufficiently detailed to discriminate between the two
main competing views.

In the past decade, some studies have argued that the continuous
infall of substructures in the dark matter halos of galaxies could
induce spiral patterns in disks by generating localized disturbances
that grow by swing amplification \citep{Gauth06,Purc11}.  According to
such simulations, the main agent producing transient features would be
satellite passages through the inner part of a disk.  Because the
tidal effects of the satellites are generally small \citep{DVFH}, this
process is distinct from interactions thought to be responsible for
grand-design spirals like M51.  
However, there are indications that dark matter substructures orbiting in the
inner regions of galaxy halos would be destroyed by dynamical
processes such as disk shocking, and hence would not be able to seed
the formation of spiral structure \citep{Don10}.

In what follows, we report on high resolution N-body simulations of
isolated disks in which we follow the motion of 100 million star
particles under the influence of gravity, to explore the idea that
spiral arms are seeded by density inhomogeneities orbiting within the
disk.  These inhomogeneities can be identified with fluctuations in
the distribution of gas in the interstellar medium of galaxies, such
as giant molecular clouds or, more speculatively, any massive star
clusters  embedded in the disk.  Because early
works already showed that the gravitational response of a stellar disk
to an overdensity orbiting in its plane is similar as the response of
a gaseous disk (see in particular Fig.7 of \citet{Toomre81}), we perform our analysis only on
stellar disks, for simplicity.

Our work focuses on understanding the origin of spiral arms and
differs in many respects from previous studies.  The methodology we
employ here has the potential to discern the physical processes occurring
in stellar disks at a higher level of detail than in previous simulations
of isolated galaxies. 
  
As we describe below, our results indicate that the
response of a disk to local perturbations is highly non-linear and
time-variable on galactic scales and so is not fully captured by the
linear approximations invoked in swing amplification theory or the
classic work on quasi-steady density waves.

In what follows, we describe our methodology in \S 2. In \S 3 
we present our results by showing examples in which a disk is
perturbed by inhomogeneities orbiting in its plane, and we even vary
the lifetimes of the perturbers.  Results anticipated by linear theory
are described in \S 4, whereas \S 5 is devoted to illustrating the
non-linear effects that arise in our simulations.  Finally, we 
discuss the astrophysical relevance of our
work in \S 6.

\section{Methodology}
\noindent
Our simulations were carried out with the parallel TreePM code GADGET-3
\citep[last described in][]{S05}. We only employ the
tree-based gravity solver coupled to a static external potential to
solve for the evolution of collisionless particles.

Pairwise particle interactions are softened with a spline kernel
\citep{HK89} on scales $h_s$, so that forces are strictly Newtonian
for particles separated by more than ${2 h_s}$.  The resulting force is
roughly equivalent to traditional Plummer-softening with scale length
${\epsilon \approx h_s /2.8}$. For our simulations the Plummer-equivalent gravitational
softening length set to ${\epsilon=5}$pc for stellar particles.

\subsection{Setting up initial conditions}
\noindent

The galaxies in our study consist of dark matter halos and
rotationally supported disks of stars. The parameters describing each
component are independent and the models are constructed in a manner
similar to the approach described in previous works \citep{H93, S00,
SDH05}.

\subsection{Dark Halo}

We model the dark matter mass distribution with a \citet{H0} 
profile:
\begin{equation}
\rho_{\rm{dm}}=\frac{M_{\rm{dm}}}{2\pi}\frac{a}{r(r+a)^3},
\end{equation}
\noindent
which has a cumulative mass 
distribution ${M(<r)=M_{\rm{dm}}r^2/(r+a)^2}$, where $a$
is the radial scale length and ${M_{\rm{dm}}}$ is the total halo mass
here set to  $9.5$x$10^{11}$
M$_{\odot}$.


In the past, models of disk galaxies run in isolation and used to
study the properties of spiral arms employed only a few million
particles to sample both the stellar disk and the dark matter halo.  In such
experiments, randomly-placed particles produce fluctuations in the
halo potential.  Even if the disk is initially featureless, the
Poisson noise owing to such discretization of the mass in N-body
experiments is inevitably swing amplified, producing trailing
multi-armed spiral patterns in the disk \citep{Toomre77,Fujii11,Sel12}.

\begin{figure*}
\epsscale{1.0}
\plottwo{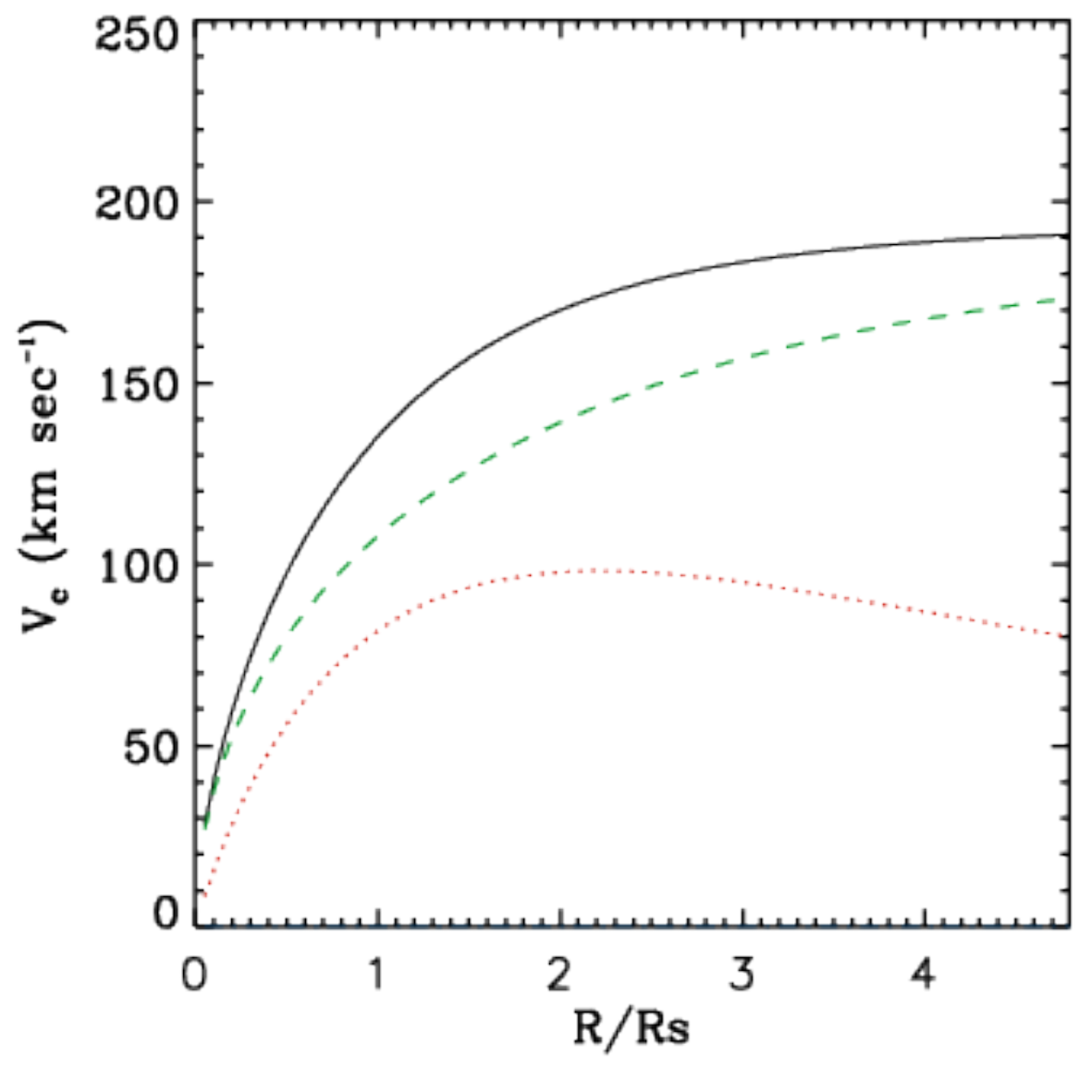}{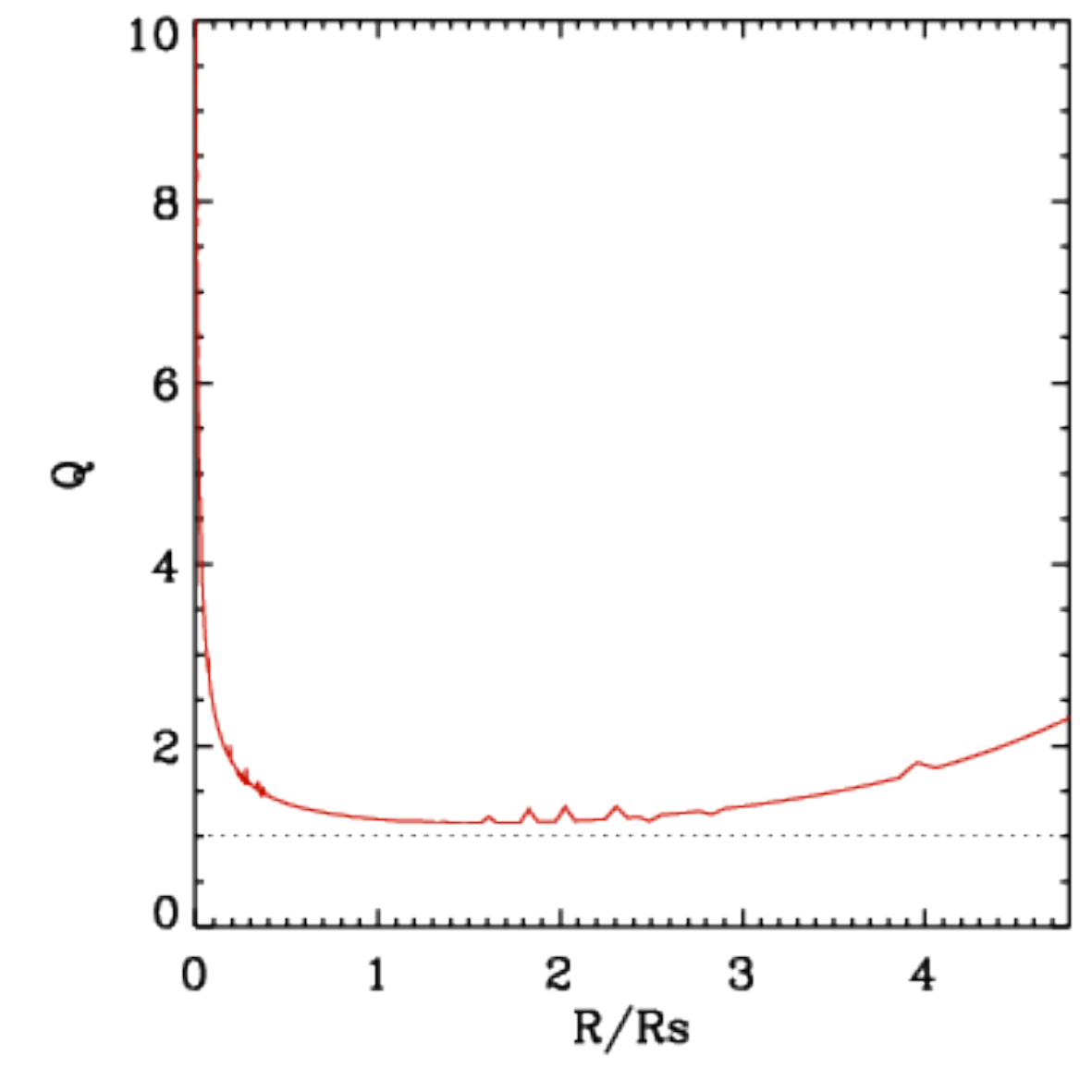}
\caption{{\it Left Panel}.Rotation curve for fiducial disk galaxy model 
(left panel). 
{\it Right Panel}. Toomre parameter $Q$ as a function of radius for the 
fiducial model. The radius is normalized to the value of the disk scale length of  3.13 kpc
in both panels.}
\label{Qvrot}
\end{figure*}

In order to suppress the development of artificial features in all the
N-body experiments that follow, we set up a live disk of stars embedded
in a rigid dark matter potential.  We employ simulations with a
sufficiently large number of particles in the disk, i.e. 100 million,
so that the disks are essentially featureless when evolved without any
perturbers acting on them. 
These simulations serve as ``controls,'' making it
possible to identify the response of the disk to imposed
perturbations.  In this manner, we will be able to separate the
sources responsible for exciting features in the disk from the stars
which react to the perturbations, unlike previous experiments in which
the stars themselves acted as perturbers, complicating the interpretation
of the experiments, as emphasized by \cite{T90}.

\subsection{Stellar Disk}

We model the stellar disk in the initial conditions with an 
exponential surface density profile
of scale length $R_s$:\\ 
\begin{equation}
\Sigma_{*}(R)=\frac{M_{*}}{2\pi R_s^2} \rm{exp}(-R/R_s),
\end{equation}
\noindent
so that the stellar disk mass is ${M_*=m_dM_{\rm{tot}}}$, where $m_d$
is the disk fraction of the total mass $M_{\rm{tot}}$ of the galaxy.
For an extremely thin disk the circular velocity of the galaxy would be:
\begin{align}\label{vc}
V_c^2(R)= & \, \frac{GM_{\rm{dm}}(<R)}{R}+\frac{2GM_*}{R_s}y^2 \times \notag \\
          & \, [I_0(y)K_0(y)-I_1(y)K_1(y)].
\end{align}
\noindent
Here, $G$ is the gravitational constant, ${y=R/(2R_s)}$,
and $I_n$ and $K_n$ are modified Bessel functions.
We specify the vertical mass distribution of the stars in the disk by
giving it the profile of an isothermal sheet with a radially constant 
scale height $z_0$. The 3D stellar density in the disk is hence given by:

\begin{equation}\label{rho}
\rho_*(R,z)=\frac{M_*}{4\pi z_0 R_s^2} \rm{sech}^2 \Big(\frac{z}{z_0}\Big) exp\Big(-\frac{R}{R_s}\Big). 
\end{equation}
\noindent
We treat $z_0$ as a free parameter that is set by
the vertical velocity dispersion of the stars in
the disk and
fix the velocity distribution of the stars such that this scale height is
self-consistently maintained in the full 3D potential of the galaxy 
\citep{SDH05}. We adopt ${z_0 = 0.1 R_s}$. 

Fig. \ref{Qvrot} shows the rotation curve for the fiducial
galaxy model used in the simulations (left panel). 
The total disk mass of the galaxy
is $1.9$x$10^{10}$M$_{\odot}$.
The total  disk fraction is fixed to 
${m_d= 0.02}$ and the scale length of the disk, $R_s$, is 3.13 kpc. 
The disk is set up so that it is stable, as
measured by the $Q$ parameter, defined for infinitely thin disks as:
\begin{equation}\label{Q}
Q=\frac{\sigma_R \kappa}{3.36 G \Sigma},
\end{equation}
\noindent
where $\sigma_R$ is the radial velocity dispersion, $\kappa$ is the epicycle frequency,
and $\Sigma$ is the stellar surface density. 
The properties of the disk are chosen so that the
$Q$ parameter is initially larger than 1 at all radii and has a minimum value of 1.3 
(right panel of Fig \ref{Qvrot}), 
implying that the disk is stable to axisymmetric instabilities  \citep{T64}.
\begin{figure*} 
\epsscale{1.1}
\plotone{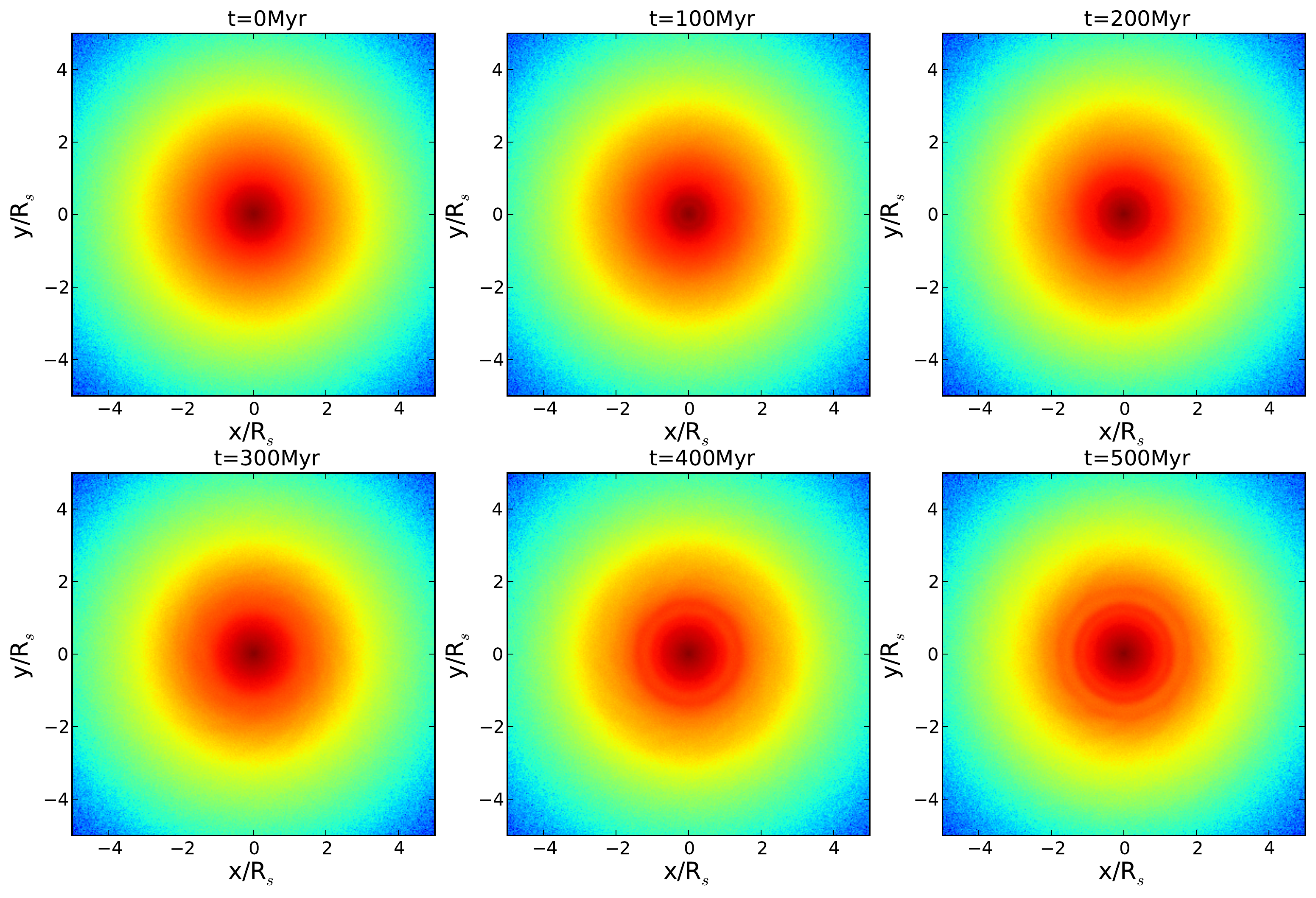} 
\caption{
Time sequence for the evolution of a live disk of stars embedded in a dark Milky Way-sized halo
run unperturbed in isolation and displayed after a few galactic years.
The disk is displayed face-on in Cartesian coordinates.  
}
\label{diskstable}
\end{figure*}

The strength of spiral features is estimated by dividing the face-on disk into concentric annuli
which are further divided into azimuthal bins.  The mass from the star particles is then
assigned to these bins, making it possible to
compute the stellar surface density ${\Sigma(R,\theta)}$.  
Fourier components of the surface density are then calculated.
We use the Fourier transform analysis to estimate the residuals of the surface density computed
by subtracting the azimuthally averaged surface density from the surface density distribution and normalizing
according to 
${Res= (\Sigma - \Sigma_{\rm{avg}})/\Sigma_{\rm{avg}}}$.

\section{Results}

\subsection{Disk Stability Test}
\noindent

We first perform an N-body experiment in which a live disk of 100
million stars as described above is embedded in a rigid potential for
the dark halo.  We run the fiducial case with no perturbers for two-three  
galactic years and verify that the disk does not develop prominent
spiral features.  The outcome of this experiment is shown in
Fig. \ref{diskstable} where a time evolution of the disk is displayed. 

\begin{figure*} 
\epsscale{1.1}
\plotone{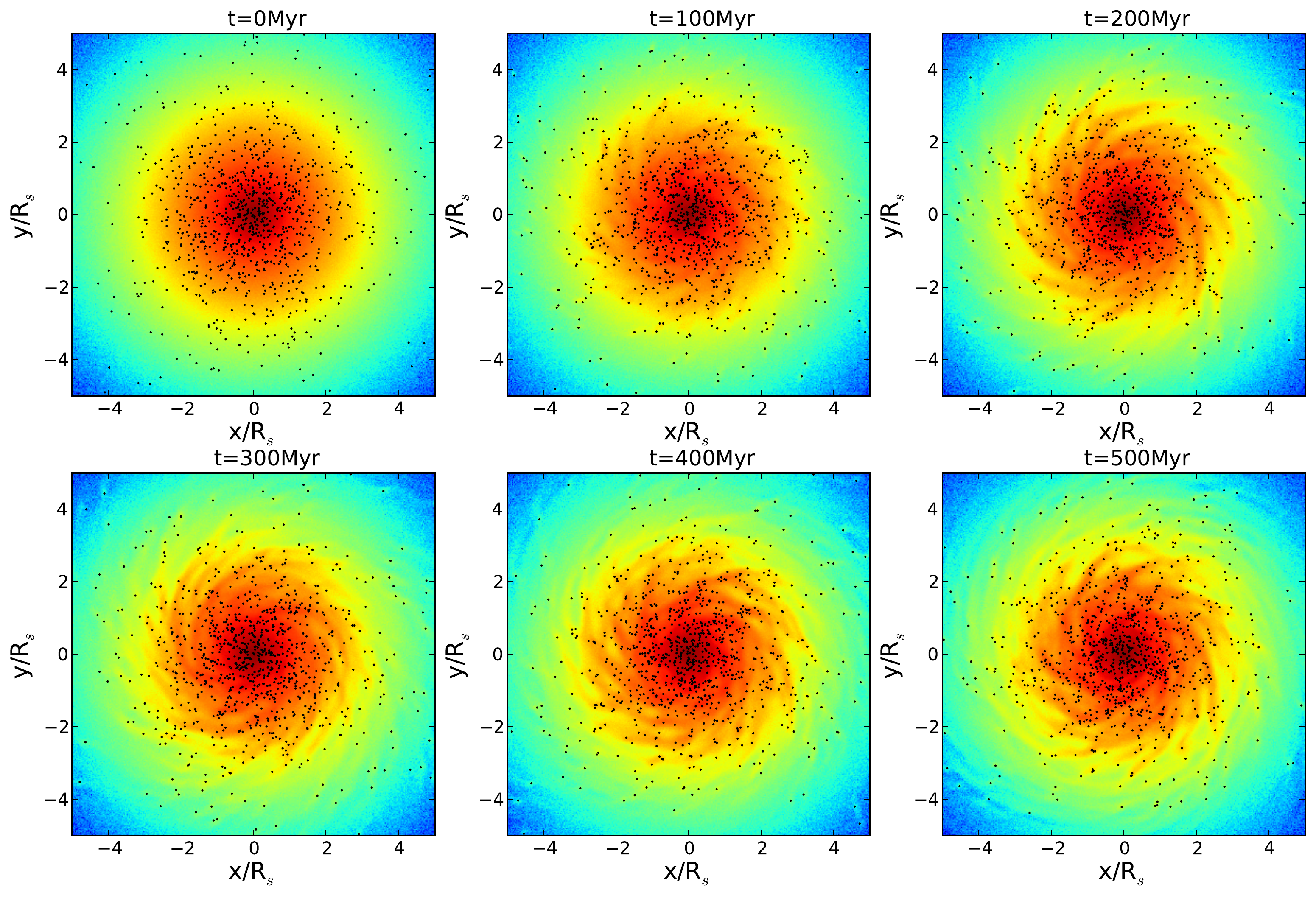}
\caption{Time sequence of an N-body experiment in which a 
self-gravitating disk of 100 million stars seen face-on  
embedded in a dark Milky Way-sized halo is run with 1,000 giant molecular clouds 
(represented with the black dots) which are 
randomly distributed within the disk and 
are assumed to be corotating on circular orbits with the disk 
stars. The giant molecular clouds act as perturbers and the live disk dynamically responds to 
the presence of these perturbers by developing features which resemble multi-armed structures 
in galaxies.  Each panel displays a region 30 kpc on a side at the times indicated.}
\label{diskpert}
\end{figure*}

\subsection{Dynamical Response of the Disk to Perturbers}

Having established that 
relatively low mass disks 
run with sufficiently large numbers of
particles do not develop prominent spiral patterns in the absence of
any perturbing influence,
we then
evolve the same galaxy models, but add $1,000$ gravitationally
softened particles 
(each with a mass of the order of a typical giant molecular
cloud: M = $9.5$x10$^{5}$ M$_{\odot}$) 
distributed with the same exponential law as the disk and
assumed to be {\it corotating} on circular orbits with the disk stars.
We note that we use a 20 times larger gravitational softening length
for these molecular cloud particles due to their larger mass compared
to the stellar particles in the disk.
These perturbers are kept moving on circular orbits in order to avoid 
any clumping that could facilitate the formation of overdense stellar regions. 

Each softened particle perturbs the motions of neighboring stars and
drives a strong {\it local} response in the disk.  The collective
influence of all the perturbers leads to the development of spiral
structures as shown in Fig. \ref{diskpert}.  These features extend
well beyond two scale lengths of the disk.  At a distance of five
scale lengths these segments tend to be more isolated and weaker.

\begin{figure*} 
\epsscale{1.1}
\plotone{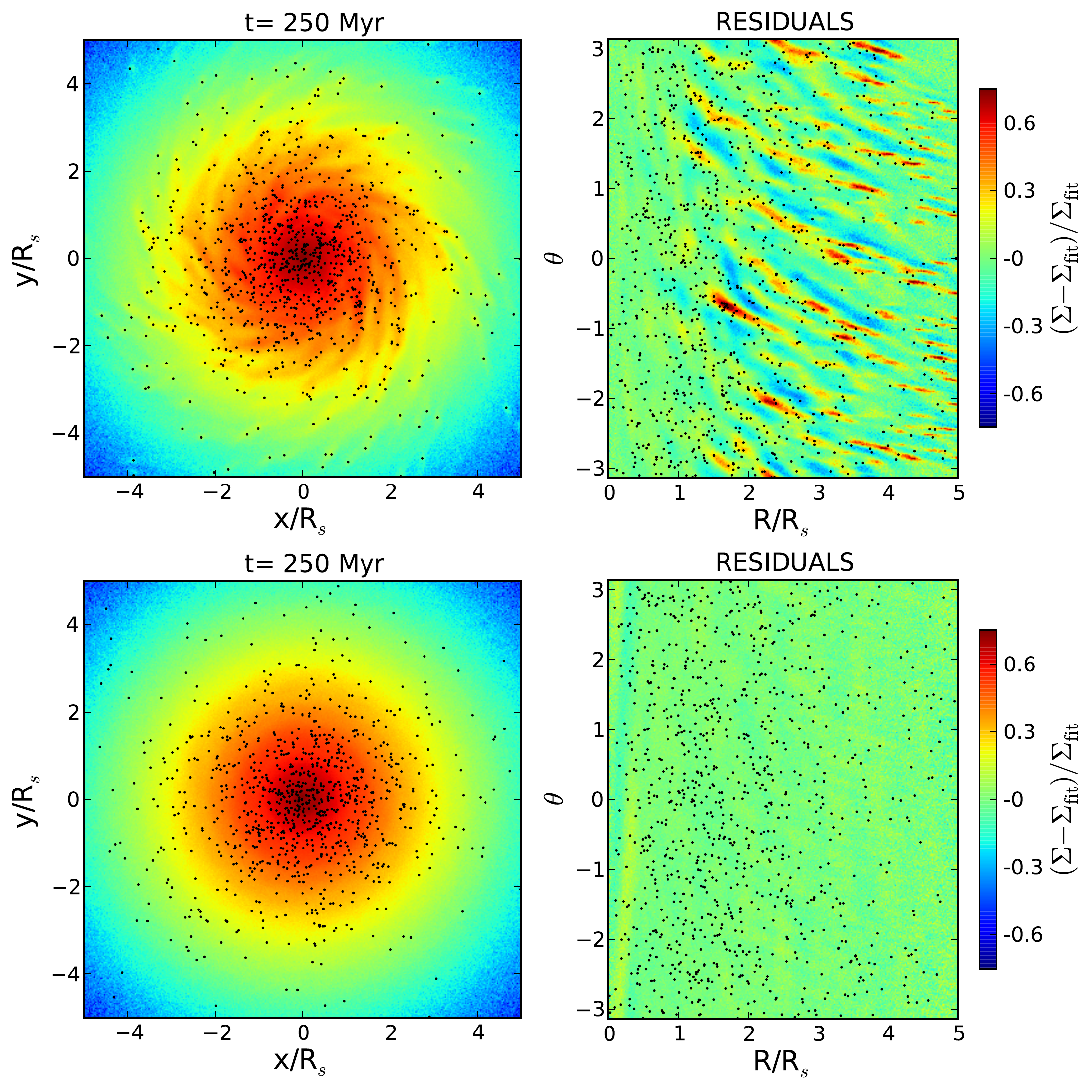}
\caption{{\it Top Panels}.A live disk of stars embedded
in a dark Milky Way-sized halo run in isolation with 1,000 giant molecular clouds  
distributed within the disk and assumed to be {\it corotating} on circular orbits with the disk. 
The disk is displayed approximately after one galactic year face-on in Cartesian coordinates (left) and 
surface density 
residuals are shown in polar coordinates (right).
{\it Bottom Panels}. The same experiment is displayed at the same time 
with the 1,000 perturbers assumed to be {\it counter-rotating} 
on circular orbits with respect to the disk.}
\label{pert}
\end{figure*}

In particular, Fig. \ref{diskpert} displays the time evolution of the
disk for approximately two-three galactic years.  Note that the dynamical
response is rapid and already after almost half of a galactic year,
multi-armed patterns appear. Fig. \ref{pert} (top panels) shows the disk and the
1,000 molecular clouds distributed within it 
after approximately one galactic year. The disk is displayed face-on in Cartesian coordinates
(left panels) and surface density residuals are shown in polar coordinates (right panels).  
In this case, interior to one disk scale length the disk is characterized by three or
four spiral arms. At two scale lengths from the galaxy center the number of spiral arms
is approximately seven with amplitudes   reaching values of 10-15\% higher than the stellar background.

We repeat the same experiment with the giant molecular clouds
counter-rotating relative to the disk stars.  As shown in
Fig. \ref{pert} (bottom panels), the disk response after approximately one galactic year is
now much weaker than for corotating perturbers displayed at the same time (top panels).

\subsection{Finite Lifetime Giant Molecular Clouds}

Our previous calculations assumed an infinite lifetime for the giant
molecular clouds.  Although the lifetimes of these systems are much
debated, giant molecular clouds in galaxies are likely short-lived.
Furthermore, giant molecular clouds in real galaxies are preferentially 
found in spiral arms, and may be created there by the converging flow of the 
interstellar medium due to its motion through the spiral potential.  
The clouds disperse after they pass through the arm, perhaps in part because 
of the diverging flow as the gas emerges from the arm, although other factors such 
as of the energy input from young stars are important also.  
This is certainly a too complicated picture to be simplified, however here
we only want to measure the dynamical response of the disk to some perturbers
with different lifetimes.   

In the following, we show the outcome of runs performed with a
finite lifetime for the giant molecular clouds.  In particular, we
show two examples: a first run adopting the fiducial disk of stars
embedded in a rigid halo potential with 1,000 giant molecular clouds
having a lifetime of 2.5x$10^{6}$ yrs and a second run with 
cloud lifetimes
of 2.5x$10^{7}$ yrs.  The finite lifetime is directly implemented in
our simulation code.  Specifically, we assign a time counter to each
giant molecular cloud particle.  Once this counter exceeds the
predefined lifetime, this particle is converted to a regular
stellar particle by changing its mass and gravitational softening
length.  We then randomly select another stellar particle in the disk
and turn it into a giant molecular cloud particle also by changing
the mass and gravitational softening length accordingly.  To avoid the
sudden destruction of all giant molecular cloud particles at the same
time, we generate the initial set of giant molecular cloud particles
with a uniform age distribution in the range of the selected lifetime.

\begin{figure*}
\epsscale{1.1}
\plotone{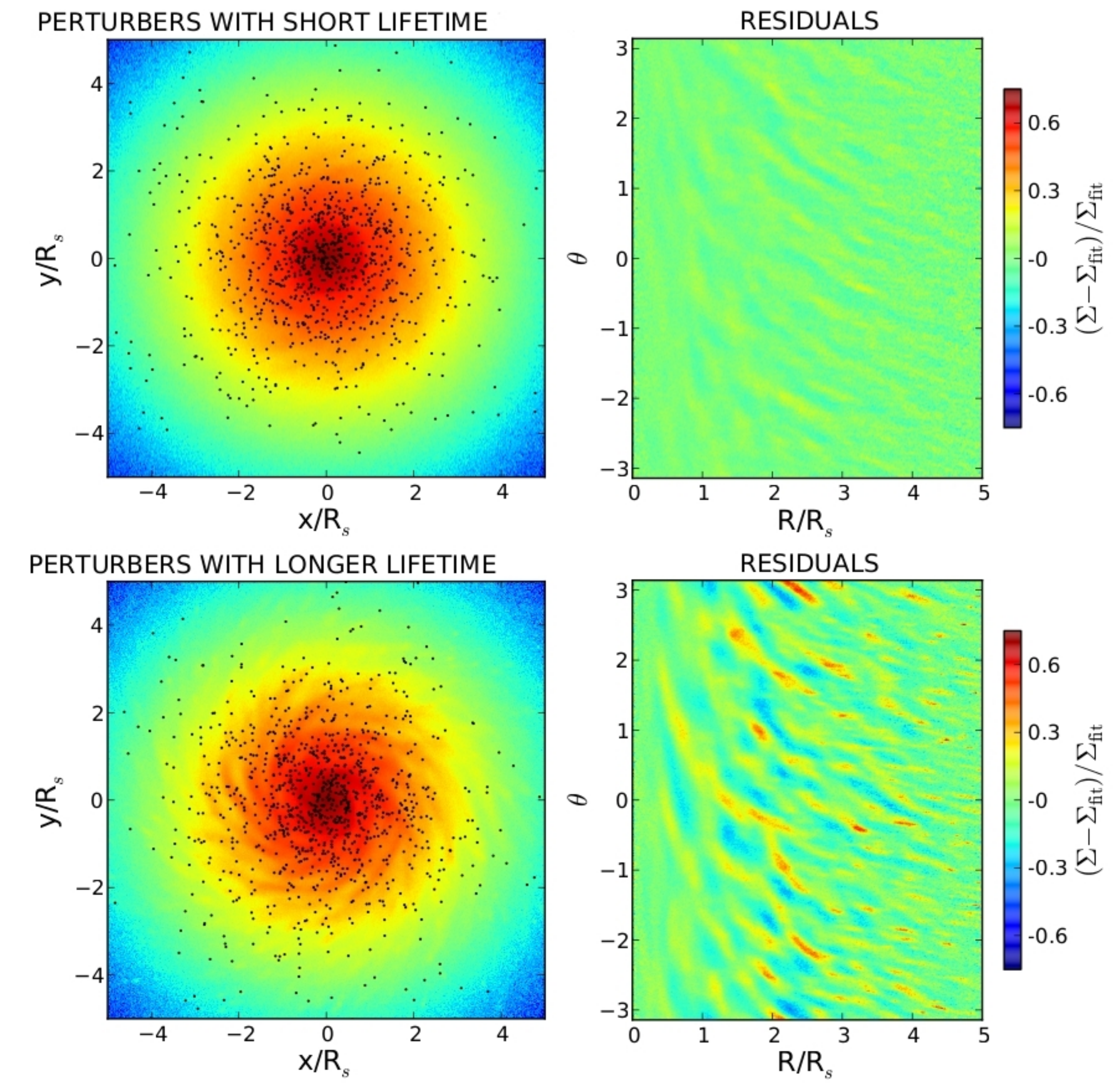}
\caption{{\it Left Panels}. The outcome of the N-body experiment where a live disk of stars embedded
in a dark Milky Way-sized halo run with  1,000  giant molecular clouds   
with a finite lifetime of 2.5x$10^{6}$ yrs (top) and 2.5x$10^{7}$ yrs (bottom). 
{\it Right Panels}. The surface density 
residuals are shown in polar coordinates.}
\label{diskmcshort}
\end{figure*}

The top panels of Fig. \ref{diskmcshort} show the outcome of the run with the giant
molecular clouds having a lifetime of 2.5x$10^6$ yrs.  This lifetime
is too short to trigger any resonance between the shear flow and
epicyclic vibration, hence the forced response that leads to the
formation of the arms does not occur. The opposite situation is shown
in Fig. \ref{diskmcshort} (bottom panels) where the giant molecular clouds are assumed
to live for 2.5x$10^{7}$ yrs.  Note that this lifetime is sufficient
to trigger the formation of multi-armed spiral patterns.  The orbital
period for a typical giant molecular cloud in a Milky-Way sized galaxy
is ${\approx 10^8}$ yrs, thus our calculations show that a lifetime of a
quarter of the orbital time is sufficient to trigger a strong response in
a disk and the subsequent development of spiral arms.  The argument can
be now inverted: in order to form realistic spiral arms in disk
galaxies, the giant molecular clouds have to survive for at least
10$^7$ yrs, which is in accord with observational estimates.

\section{Results anticipated by linear theory}

\subsection{Material Arms or Density Waves?}
\begin{figure*}
\epsscale{1.1}
\plotone{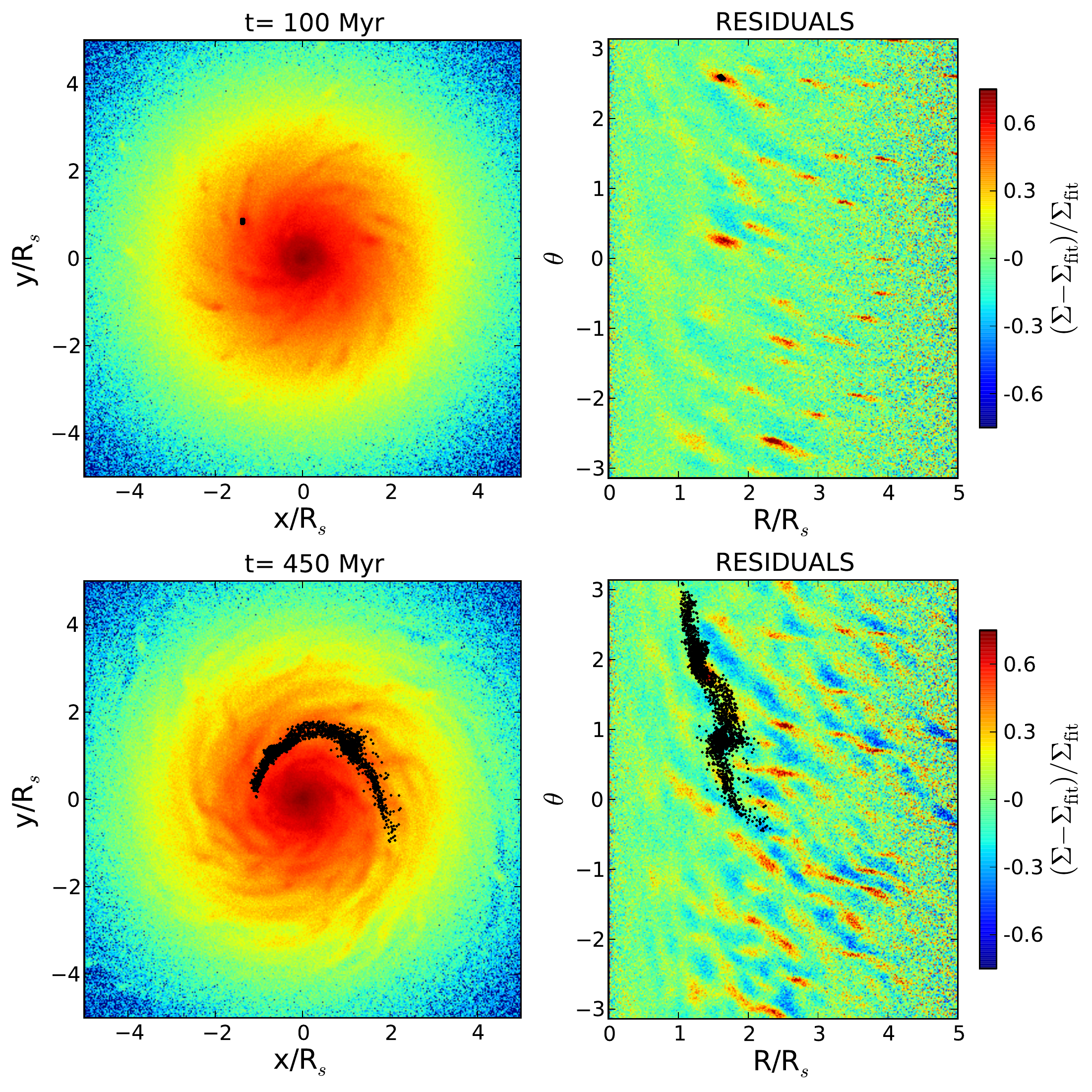}
\caption{{\it Top Panels}.
A patch of stars is identified in the overdense region of the arm and colored in black at the time when
the arms are fully developed
(100 Myrs). {\it Bottom Panels}. 
The stars initially identified in the patch are followed forward in time and displayed after less than two  galactic years.}
\label{waves}
\end{figure*}

\noindent
A long-standing controversy over the nature of spiral arms is whether
they correspond to density enhancements in the background stellar
distribution (density waves), or are made up of stars that always
remain in the arm and are just more concentrated than the stars
outside the arm (material arms).  In the early studies, the arms were
assumed to be density waves, because if they were material they would
quickly wind up as the galaxy rotates.  Thus, both the swing
amplification and the static density wave theories argue that the arms
are overdense regions of the disk moving around at a different speed
relative to the stars themselves.  Stars thus continuously move in and
out of the spiral arms.  However, recent investigations using
numerical simulations of stellar disks have challenged this claim and
and argue that the arms might be material structures \citep{Grand}.

To investigate this in the context of our simulations, we identify a
patch of stars along the arm in the stellar disk after the arms are
fully developed (after 100 Myrs) as displayed in top panels of
Fig. \ref{waves}, where the patch is colored in black.  Then, we follow
the positions of the stars originally in the patch forward in time and
display the outcome after two galactic years (bottom panels).  We note
that the stars initially in the patch spread out, confirming that the
spiral patterns in our simulations are density waves and not material
structures.  This is shown in polar coordinates in the
bottom panel of Fig. \ref{waves} where it is clear that the patch is
being sheared out by differential rotation and the pitch angle of the
patch differs significantly from that of the spiral features.

\subsection{Forced-Wake Theory} 

The different response of the disk to the direction of the motion of
the perturbers within the disk, as shown in
Fig. \ref{pert},
suggests that the physical process
behind spiral arm formation in our simulated galaxies is 
initially, at least, wake-making, which operates through a combination of
three ingredients: the shearing flow, small-scale epicyclic motions, and
the disk self-gravity \citep{JT66}.
We briefly
review the main concept.

Consider stars drifting in
the flow at slightly larger radii than a perturber which is
corotating with the disk.  The drifting stars will
feel a small force owing to the presence of the
perturber, exciting epicyclic motion in the stellar orbits.
At the same time, the inwards and outwards motions of the
stars lead to density enhancements and deficits relative to
the stellar background.  These enhancements can swing from
leading to trailing owing to shear in the disk.  The sense of
this shearing motion is identical to that of the retrograde
epicyclic motion of individual stars.
Because of this match it
is possible for a swinging pattern to resonate with the
epicyclic stellar motions as the shear causes it to
become a trailing pattern.  This temporary match allows
the feature to be amplified by
self-gravity as the stars linger within the perturbation
\citep{JT66}.

\subsection{The Critical Role of Self-gravity as Amplifier}

\citet{JT66} showed that swing amplification results from
a cooperative effect between shear, epicyclic 
motions, and self-gravity.  In order to verify the importance
of self-gravity in the outcomes of our simulations,
we performed
an N-body simulation of a disk of stars treated as test particles
embedded in a dark halo along with 1,000 perturbers moving on 
circular orbits within the disk (for details of the
set-up, see Appendix).

\begin{figure*}
\epsscale{1.1}
\plotone{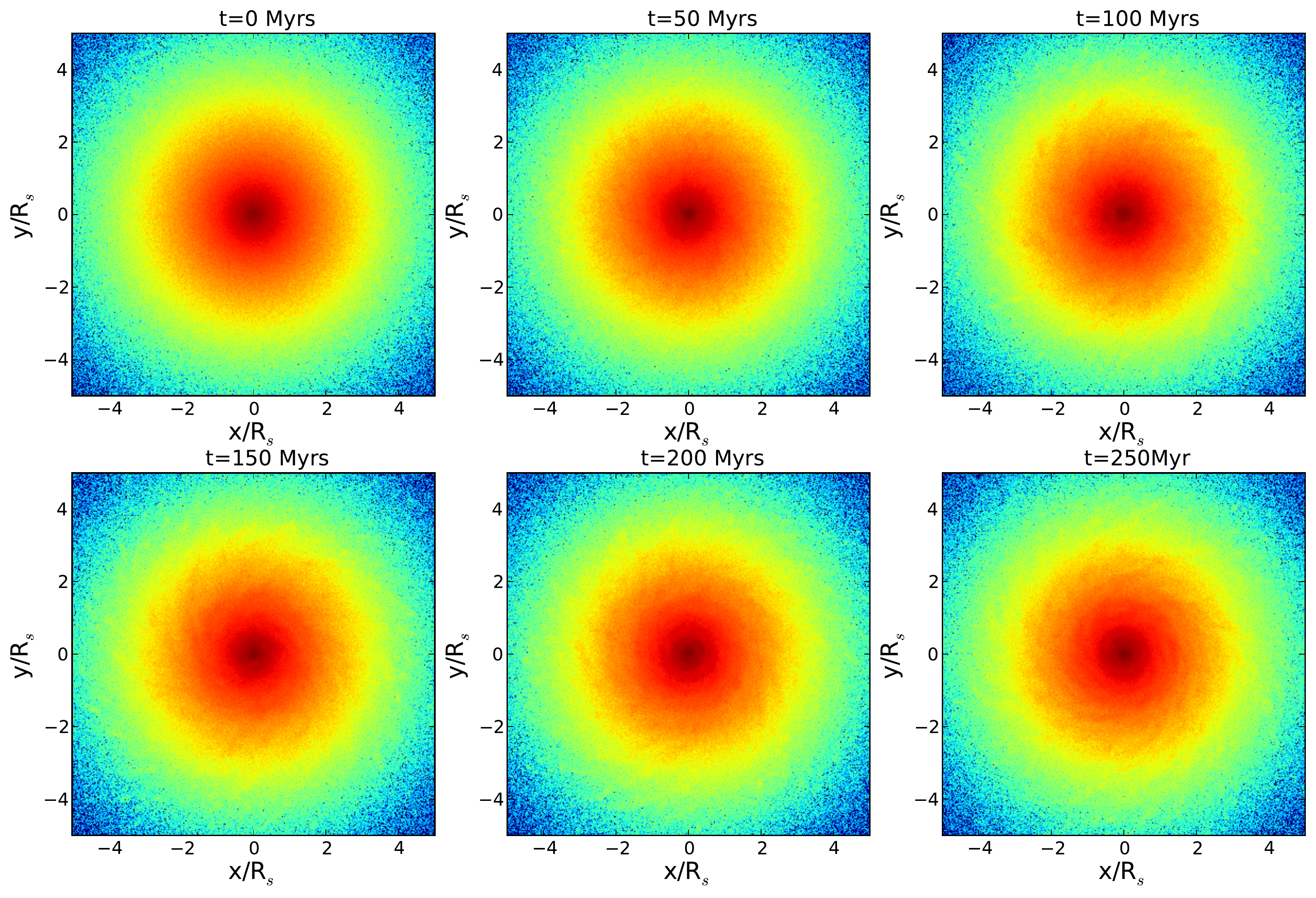}
\caption{Time evolution of a disk perturbed by 1,000 inhomogeneities
co-rotating with the stars, but in which the disk self-gravity is
not included.}
\label{noself}
\end{figure*}

Fig. \ref{noself} shows a time sequence from a simulation in which the
disk is perturbed by density inhomogeneities, but the self-gravity of
the disk is ignored.  The influence of the disk self-gravity is clear.
In the absence of self-gravity, wakes still form around the
perturbers, but the wakes have different shapes and are weaker than
when self-gravity is included, as demonstrated in Fig. \ref{noself}.
A harmonic analysis shows that the magnitude of each Fourier
coefficient is greatly reduced if disk
self-gravity is ignored, as expected on the basis of similar tests
performed by \cite{JT66}.  (In particular, compare Figs. 7 and
12 in \cite{JT66}.)
Thus, self-gravity acts as a communicator
between the stars, amplifying wakes around each perturber and allowing
different wakes to connect to produce long-range patterns, as in
Fig. \ref{diskpert}.

\subsection{Collective Amplification Regime}

Swing amplification is the linear gravitational
response of a disk to an overdensity orbiting in its plane.  In a
stable disk, pressure-like or mixing effects of random motions avoid
clumping and normally disrupt an overdense region before it has time
to collapse.  Because of the similarity between shear and
epicyclic vibrations
however, individual stars remain within shearing density
enhancements
for a substantial
fraction of the epicyclic frequency $\kappa$, which is given by
\begin{equation}
\label{kepyc}
\kappa^2=R \frac{\rm{d}\Omega^2}{\rm{d}R}+4 \Omega^2,
\end{equation}
\noindent
where ${\Omega=V_c/R}$ is the angular frequency,
$R$ is the radius, and $V_c$ is the total circular velocity.
For the idealized case of a Mestel disk, with $V_c$ constant, 
\cite{Toomre81} found that 
the gain of the swing amplifier 
depends on the value of $Q$ and the ratio of the wavelength to a critical
value ${X=\lambda/\lambda_{crit}}$ where ${\lambda_{crit}=4\pi G \Sigma \kappa^{-2}}$ 
is the shortest wavelength stabilized by rotation alone.
For this idealized situation,
the amplification factor
peaks at ${X \sim 1.5}$, reaching values in excess of 100 for cold disks.
Swing amplification is negligible for ${X \ge 3}$. 
Indeed perturbations with such long wavelengths are stabilized by
the angular momentum of the stars.
Note that in N-body experiments, particle noise creates a spectrum of perturbations of all sizes and shapes, including both leading 
and trailing spirals. 
As the differential rotation of the disk shears leading spirals into trailing configurations, swing 
amplification boosts the amplitudes of those which have wavelengths of order of ${\lambda \sim 1.5 \lambda_{crit}}$ thereby creating a 
multi-armed pattern of trailing spirals with a characteristic spacing that explains the results illustrated in 
Fig. \ref{diskpert} (see Appendix for an
expression giving the amplification parameter for a general disk galaxy).

A naive expectation from \citet{JT66}  is that each inhomogeneity
in the disk should generate a wake, forming an initial leading feature
which is amplified owing to the differential rotation of the disk to
become a stronger trailing wave.  However, in our simulations the
number of arms is not determined by the number of perturbers.\footnote{
It should be noticed that in linear theory, there is fundamentally no difference, 
whether or not the perturbers are distinct from the supporting medium \citep{T90}.}
Instead, each perturber produces a segment of an arm and then these
segments are joined at kinks.  {\it Thus, there is a collective
process that occurs between the individual wakes that connects them
together to produce a pattern that resembles the features in
multi-armed spiral galaxies.}

According to swing amplification theory, the most strongly amplified
features are those which have wavelength ${\sim \lambda_{crit}}$.  If in
some region the perturbers are packed together more tightly than this,
they will effectively act as an individual perturber producing a
single response with an extent ${\sim \lambda_{crit}}$.  In regions
where the perturbers are separated by distances much larger than
${\lambda_{crit}}$, they are effectively isolated, and we would expect
to see individual, disconnected wakes around each perturber.  However,
in the intermediate regime, when the density of the perturbers is such
that they can each produce wakes that interact with one another,
the wakes will connect owing to self-gravity, joining together to
produce patterns that can cover an entire disk.  This has the
effect of selecting out particular global structures that
extend throughout a disk and the resulting number of
arms is governed by the structural properties of a galaxy
through ${\lambda_{crit}}$.

\subsection{Determination of the Number of Arms}

So far we have presented  models of disks of such mass that when they
are perturbed they develop features characteristic of multi-armed
spiral galaxies.  We have verified that the number of arms does not
depend on the number of perturbers.  However, the response
does depend in detail on the mass distribution of the galaxy and the
extent to which the disk is self-gravitating.

\begin{figure*}
\epsscale{1.1}
\plotone{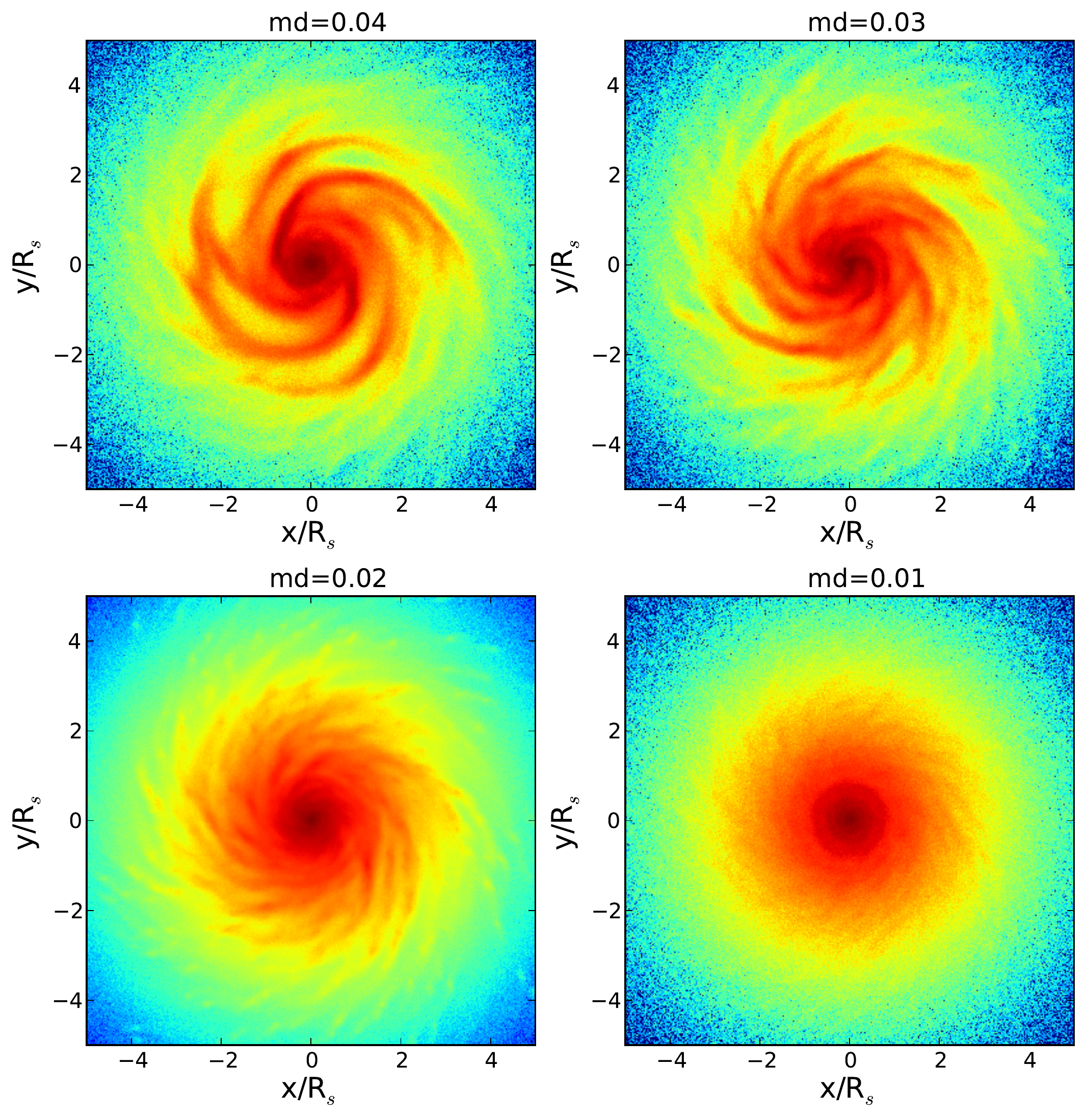}
\caption{Sequence of models where we varied the contribution of the disk to the total mass 
distribution in the rotation curve of the galaxy. We fix the total mass of the galaxy 
and the mass distribution of the dark halo,  but increase the disk mass fraction $m_d$  
from a value of 0.01 to the following values: 0.02, 0.03 and 0.04.  
Increasing the disk fraction decreases the number of arms. Models are displayed after
one galactic year.}
\label{arms}
\end{figure*}

In a set of experiments shown in Fig. \ref{arms}, we altered the
contribution of the disk to the total mass of the galaxy.  We kept the
total mass of the galaxy and the mass distribution of the dark halo as
previously but we ran simulations where we increased the disk fraction, $m_d$,   
from 0.01 to 0.02, 0.03 and 0.04. Once again, we
first ran these models without perturbers orbiting within the disk for one galactic year and
verified that the disks do not develop prominent patterns.  We
subsequently introduced in the initial conditions 1,000 giant
molecular clouds corotating with the disk stars.  By changing the disk
mass these models were designed to vary the critical length scale
parameter ${\lambda_{crit} = 4\pi^2 G\Sigma/\kappa^2}$ and lead to a
different spiral morphology (see Fig.1 in \citet{CF85,ELN82}), from the multi-armed
features typical of galaxies like NGC 7217 obtained for low-mass
disks, to those with a few prominent arms as in M101 or the Milky
Way.  As anticipated based on the above arguments, the results 
shown in Fig. \ref{arms} demonstrate that the spiral 
morphology in our simulated disks is indeed determined by the
structural properties of the galaxy through $\lambda_{crit}$.

\section{Results not anticipated by linear theory}

\subsection{Self-perpetuating spiral arms}

Theories based on a linear approximation describing the local dynamic
response, e.g. a disk patch, predict that the emerging spiral patterns
should be a superposition of swinging wavelets, shearing with the
general flow, but exhibiting transient growth as they swing from
leading to trailing.  If the perturbation responsible for exciting a
wakelet is removed, the linear theory predicts that the wakelet will
decay, revealing its transient nature \citep{TK91}.  In order to test
the validity of the linear theory and to determine if spiral
structures arise collectively through a local response of the disk, we
stacked the giant molecular clouds at a fixed radius in the disk,
superposing their associated wakes.  The outcome is surprising, and we
find that the response is highly non-linear in the sense that the
wakes that initially formed around the perturbers, e.g. the giant
molecular clouds, depart from them after only about a tenth of a galactic
year and then become new perturbers that continuously excite the disk.

\begin{figure*}
\epsscale{1.1}
\plotone{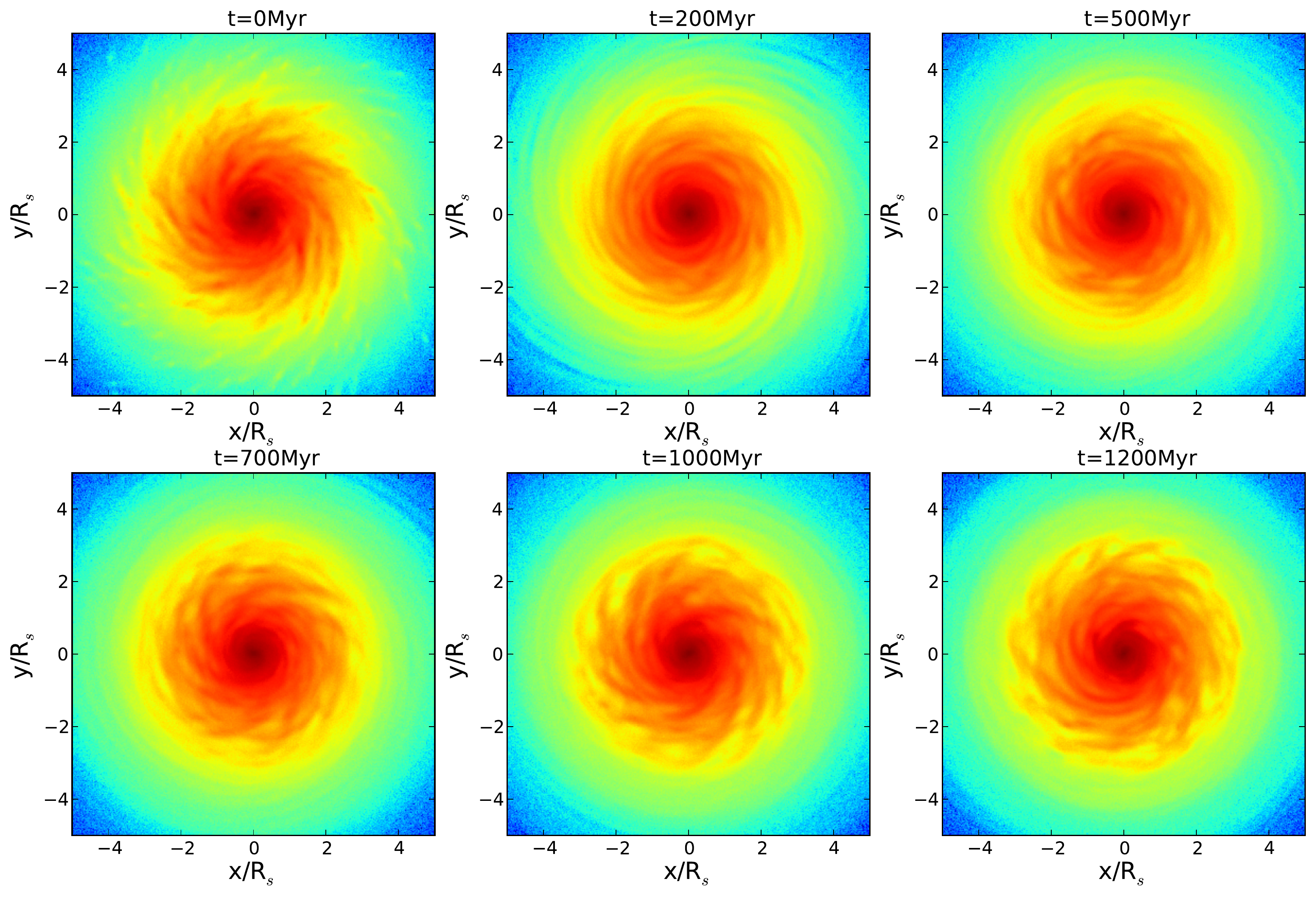}
\caption{
Time sequence for the evolution of spiral arms once the giant molecular clouds are 
removed from the disk. The left upper panel shows the stellar disk the time
at which the perturbers, e.g. giant molecular clouds, are removed. We define this initial time 
as set to t=0. The subsequent time evolution of the spiral arms is shown in the other panels
according to the initial time definition.
 Each panel displays a region 30 kpc on a side at 
the times indicated.}
\label{switchoffgmc}
\end{figure*}

In fact, this is apparent in the top right panel of Fig. \ref{pert} where it
can be seen that at a fixed radius between say 2 and 4 disk
scale-length, the wakelets have different shapes and cannot be
superposed.  Furthermore, the perturbers (the black filled circles)
are associated with their each individual wakelets, as predicted by
linear swing amplification theory, only at distances beyond 4 disk
scale-lengths where the self-gravity of the disk is lower and
the individual perturbers are well-separated.  As described below, we
find that between two and four disk scale-lengths the arms are
reinforced owing to non-linear effects and the wakelets become
sufficiently overdense that they decouple from the original perturbers
and then act as perturbers themselves.
 
To demonstrate this unexpected outcome and its radical consequences,
we remove the original perturbers from the disk by replacing the giant
molecular clouds with an equal mass in stars, and continue to evolve
the disk, as illustrated in Fig. \ref{switchoffgmc}.  The patterns do
not disappear, contrary to the expectations of linear theory, and are
instead long-lived.  Because of the self-gravity of the disk, the
wakes imprinted by the original perturbers become mildly
self-gravitating and are of sufficiently high density to serve as
perturbers themselves, thereby making the spiral features
self-perpetuating.  We have verified that the ensuing spiral patterns
survive for at least another eight galactic years.

We emphasize that in our simulations, the spirals are
self-perpetuating in a manner completely distinct from the classical
density wave theory where the patterns are static or slowly rising in
amplitude with time and are rigidly rotating \citep{Lin64,BL96}.  Indeed, in our
simulations the patterns are not {\it global} but they appear as local segments
connecting together and fluctuating in amplitude with time.  
The evolution of the resulting spiral structure is characterized by a
balance between shear and self-gravity.  Locally, shear tends to
stretch the waves, breaking the arms, whereas in regions where the
self-gravity dominates, the pattern is overdense and generates the
segments making up the arms.  This local balance between shear and
self-gravity gives the visual impression that the spiral structures
are global patterns, but in reality they are only segments, produced
by local under-dense and over-dense regions.

An alternative explanation for our results involves the
possibility that such dynamical systems support actual coherent
unstable modes, and that the features observed after the perturbers
are removed are manifestations of these unstable modes.  Thus the
rapidly varying spirals seen in our simulations would result from the
superposition of a few long-lived waves (for a detailed study, see
\citet{Sel12}).

We generated spectrograms to test this interpretation, which supported
at some level this possibility.  However, our experiments were not run
for a sufficiently long time to be decisive on this point.  We will
explore this alternative explanation with tests to discriminate
between the two interpretations in a forthcoming paper.

Our results naturally predict that the pattern speed of spiral arms in
late-type spiral galaxies should not be constant but should vary with
distance from the center of the galaxy.  This has already been claimed
for patterns speeds estimated for NGC 5164 and NGC 628
\citep{Foy10,Meidt} and M101 \citep{Foy11}.  However, whereas these 
works conclude that this supports the interpretation that spiral
arms are short-lived, our simulations show that the features can
be long-lived as various local segments break and then later
reconnect with other pieces of spiral arms, resulting in
long-lived structures.

Note that we have focused on 
the idealized case of a rather low disk mass. 
These systems are stable over
long periods of time and do not develop bars or arms if unperturbed.
If we consider disks that are more strongly self-gravitating, by, for
example, increasing the ratio of the mass of the disk to the total
mass of the galaxy, $m_d=0.04$,   
the simulations favor the formation of four spiral arms as displayed
at the top left panel of Fig. \ref{arms}.  In that event,
$\lambda_{crit}$ is larger, so that longer wavelength disturbances are
most strongly amplified.
By examining the rotation curves of our galaxy models,
we also find that those with more massive disks have
lower shear rates where the bulk of their mass is located,
so individual wakes can remain connected over longer distances
without breaking apart owing to shear, yielding
patterns with fewer global arms. Indeed the number of arms does not depend 
on the number of perturbers but on the structural properties of the galaxies
in a way that by changing the rotation curve of the galaxy also the number
of arms varies accordingly.

In a further set of experiments,
 we ran simulations of disks after we systematically
reduced the time at which the 1,000 perturbers, e.g. giant molecular
clouds, were removed and estimated
the minimum time the
perturbers need to corotate with the disk to trigger the formation of spiral arms
that persists even after being removed.  We find that if we remove the 1,000
perturbers after 15 Myrs, this time is sufficient to trigger the
process that leads later to the formation of long-lived, self-perpetuating
patterns, similar to the estimates we made earlier for the required
lifetimes of the original perturbers.

\subsection{Non-linear effects}

\begin{figure*}
\epsscale{1.1}
\plotone{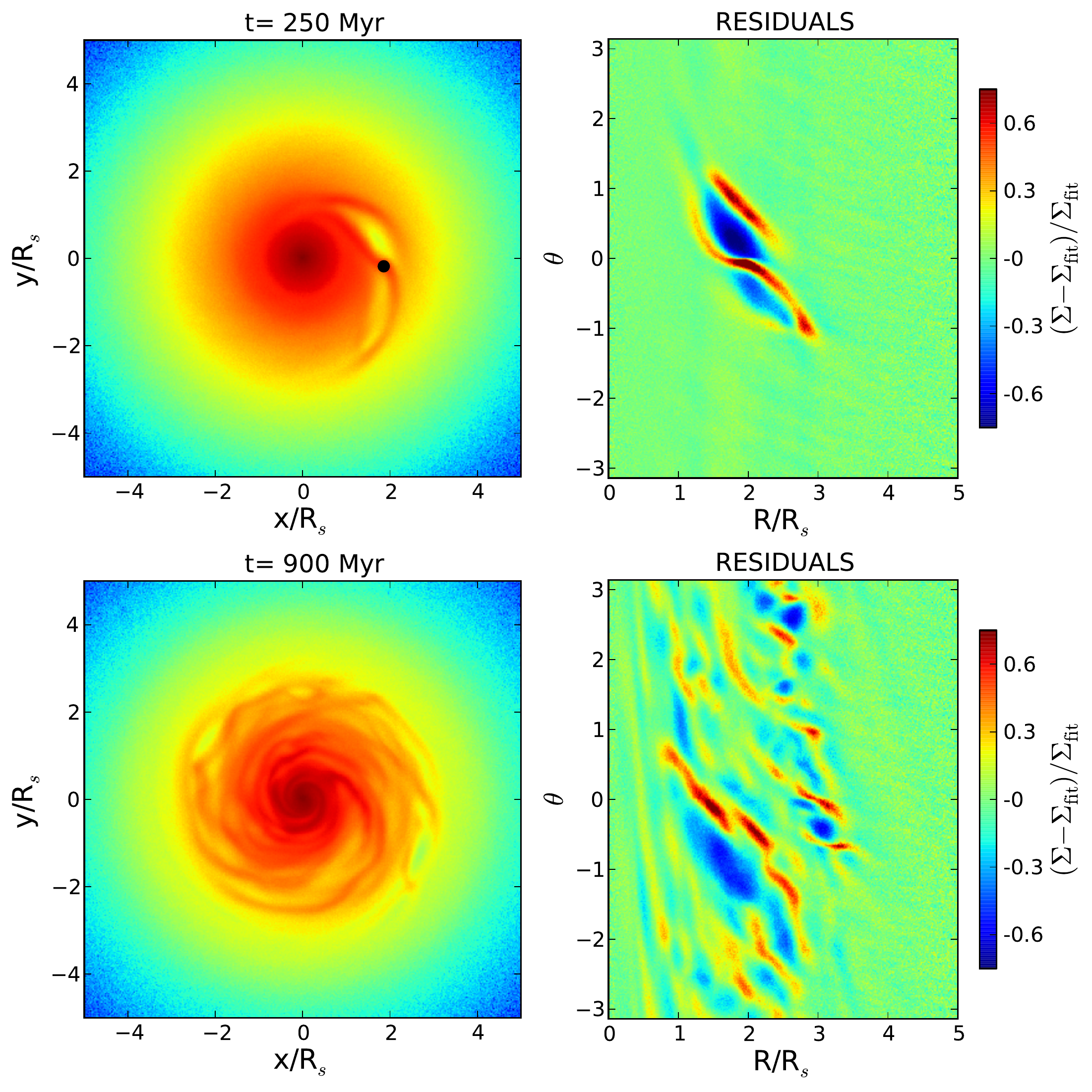}
\caption{{\it Top Panels}: A stellar disk responding to a single perturber with mass of 10$^{7}$ M$_{\odot}$,
which is removed after one galactic year. {\it Bottom Panels}: 
Surface density is shown 4 galactic years after the perturber is removed confirming 
that the patterns survive for several rotation periods.  
The disk is displayed face-on in Cartesian coordinates (left panels) and 
surface density 
residuals are shown in polar coordinates (right panels).}
\label{SwitchM7}
\end{figure*}

To demonstrate explicitly the possible
importance of non-linear dynamics in spiral structure
formation, the top panel of Fig. \ref{SwitchM7} shows the case of a galactic disk
influenced by a single perturber with a  mass of $10^7$ M$_{\odot}$ kept
cycling at 2 disk scale length, corotating with the disk,
and then being removed after one galactic year.\footnote{The perturber is forced to cycle on 
a circular orbit so that it is not affected by dynamical friction.}  The image displays
the disk when the perturber is still corotating with it.  Initially, a
wakelet is generated around the perturber in a manner consistent with
the linear theory.  However, non-linear effects are evident in the
gravitational response of the disk to the perturbation and result in
the  development of additional underdense and overdense regions that
are clearly visible especially in the polar coordinate plot (right top 
panel of Fig. \ref{SwitchM7}).  These features, which act as perturbers, are responsible 
for keeping alive the spiral activity and are not expected based on
the linear analysis. These findings indicate that our simulations are probing an
entirely new regime in which the linear theory is no longer fully
applicable.

Once the non-linear features begin to develop, we remove the perturber
as shown at the bottom panel of Fig. \ref{SwitchM7} and find that the wakelets do not
disappear, as predicted by the linear theory.  Indeed, the overdense
and underdense 
regions act themselves as new perturbers, exciting a further response
in the disk, leading to a self-perpetuating pattern in which
additional unexpected features such as holes and wrapping arms in the
outer regions are produced within four galactic years, as shown in
Fig. \ref{SwitchM7}.  The expectations of linear theory are that the
disk should relax after the perturbation is removed and that the
spirals should gradually disappear, but we find instead that the
patterns survive for many rotation periods.

We note here that the response in the disks we simulate originates at
the co-rotation resonance between the perturbers and the surrounding
stars. Thus,
while we do not argue against the possibility that dynamical effects
such as ``groove instabilities'' \citep{Sel89} can play an important role in producing
spiral structure in actual galaxies, in our simulations, at least, the
dominant dynamical effects seem to be associated with the response at the
co-rotation resonance in the disk.

\begin{figure*}
\epsscale{1.1}
\plotone{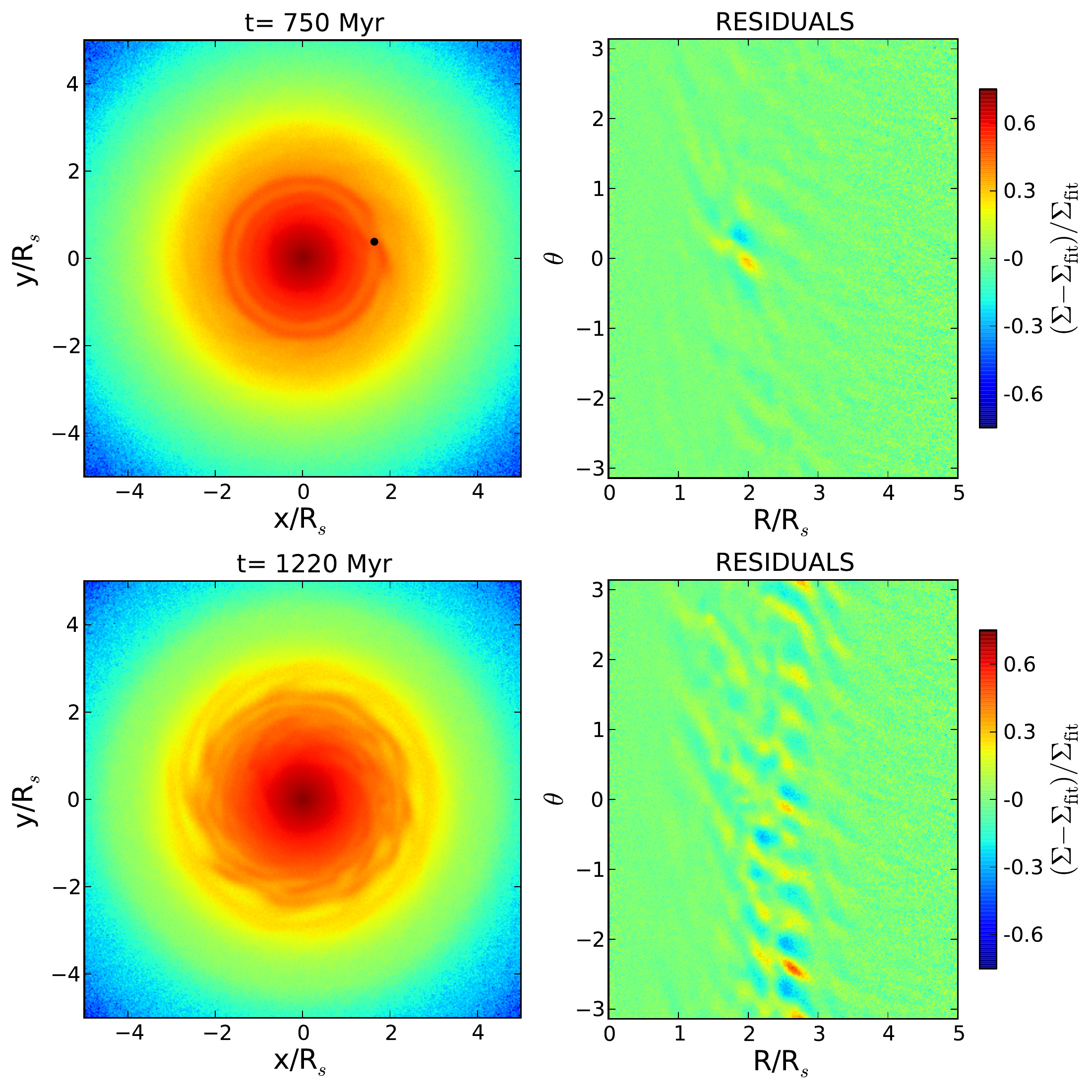}
\caption{{\it Top Panels}: A stellar disk responding to a single perturber with mass of 10$^{6}$ M$_{\odot}$,
which is forced to remain
on a circular orbit (represented by the black circle). The image displays the disk after three orbital periods 
computed at a distance of two scale lengths of the disk where the first non-linear effects
are visible. 
{\it Bottom Panels}: The patterns 
survive for several rotation periods and surface density  is shown after 2 
additional galactic years once the perturber is removed.
The disk is displayed face-on in Cartesian coordinates (left panels) and 
surface density residuals are shown in polar coordinates (right panels).}
\label{OneGMCM6}
\end{figure*}

Next, we performed the same experiment with a perturber mass reduced
by a factor of ten, e.g. typical of a giant molecular cloud or a
globular star cluster of mass $10^6$ M$_{\odot}$ kept cycling at two
scale lengths of the disk and checked how long it takes for non-linear
effects to develop.  The outcome is illustrated in Fig. \ref{OneGMCM6}
which shows that the linear approximation breaks down after roughly
3 galactic years and holes and winding arms appear following an
additional 2 galactic years once the perturber is removed.  This
suggests that while reducing the mass of the perturber lengthens the
time for non-linear effects to become important, they nevertheless
still develop.  Note that the amplitude of the arms (which are the
overdense regions) and the depth of the holes (the underdense regions)
are also reduced because they are sensitively dependent on the mass of
the perturber.

\section{PREDICTIONS}

It seems clear from our results that a modest overdensity
{\it corotating} with a galactic disk for a small
fraction of a galactic year can trigger the formation of spiral
structure that will take over and become self-perpetuating for at
least several billion years.  Intriguingly, observations of nearby
disk galaxies suggest that when {\it counter-rotating} stellar or gas
disks are present in galaxies there is no evidence of spiral structure
\citep{Ber,Cocc}, which is as expected, based on our simulations.

Furthermore, it has been argued that spiral structures are transient
features that need to be re-generated by continuously adding new
perturbers in the form of giant molecular clouds or clumpy gas that
maintains ragged arms \citep{SelCarl84}. 
Unlike simulation results of decades ago,
which were compromised by limited numerical resolution, our experiments
show that disks do not 
necessarily thicken because of the presence of spiral arms
and giant molecular clouds, and that the spiral structure can persist
for timescales similar to the ages of actual galactic disks.
Moreover, although new gas accreted can excite young arms, there
should be old stellar arms present in galaxies and fresh gas is not
required to maintain spiral patterns over long timescales.  This means
that, in principle, early-type galaxies which are now gas-poor may
still exhibit weak spiral features.  IC 3328, an early-type dwarf
galaxy with spiral structure \citep{Lisker} or NGC1533
\citep{DGraaf07} may be such cases.
 
Ultimately, other predictions of our models include definite
quantitative statements regarding the pattern speeds and
time-variability of actual spiral features in galaxies (Nelson, D'Onghia \& Hernquist 2012, in prep.), 
and the kinematic and spatial response of the stars to these fluctuating
features.  Spiral waves in galaxy disks churn the stars and gas in a
manner that largely preserves the overall angular momentum
distribution and leads to little increase in random motion.  However,
recent works \citep{SB02} claim that changes in the angular momenta of
individual stars are typically concentrated around the co-rotation
radius for an individual spiral wave, but given the supposed transient
nature of the waves with a wide range of pattern speeds developing in
rapid succession, the entire disk should be affected.  Our models
indicate that spiral structures are not transient features, but
eventually locally break and subsequently reconnect as a balance
between the shear that stirs the arm and the self-gravity that
tends to reconnect the pattern.  This calls for a better understanding
for how stars like the Sun can migrate if the nature and the evolution
of the waves is different \citep{Rok08,Minc11}

We anticipate that there should be a correlation between the amplitude
of the arms and the {\it Q} parameter that defines how dynamically hot
the disk is, in the sense that ``colder'' disks should, on average,
display more prominent spirals.  This follows from the fact that swing
amplification, which is still central to our interpretation, is most
vigorous in dynamically cool disks \citep{Toomre81}.  Ongoing and
planned observations \citep{Elm} will enable tests of our 
modeling in
the near future, which should shed further light on the
still-mysterious nature of spiral structure.

\section{CONCLUSIONS}

In this paper, we have examined the possibility that spiral structure
in disk galaxies originates through the dynamical response of
self-gravitating, shearing disks to local density perturbations.
While our analysis has much in common with earlier studies of 
spiral structure formation, it differs in the following respects.

\begin{itemize}
\item Our N-body simulations show that spiral patterns can be 
described by density waves but the response of the disk 
to local perturbations can be highly non-linear and time-variable.
Our work indicates limitations of
two traditional
theories that tried to explain arm formation: the theory
of arms as static density waves \citep{Lin64} and linear forced-wake
theory \citep{JT66} which were based on simplifying
assumptions.
In particular, our results are inconsistent with the idea that spirals
are static density waves because the arms in our simulations 
fluctuate in time.
Moreover, our results are different from what is expected
based on forced-wake theory
because we show that the arms can be long-lived and are not
necessarily transients even 
when the original perturbations responsible for
driving the patterns are removed.

\item The spiral patterns in our simulations
are not global as predicted by the classical static 
density wave theory but locally they appear to fluctuate in amplitude with time.  
It is the local balance
between the shear that stirs waves, breaking the arms, and the self-gravity that makes
up the arms  
that gives the visual impression that the patterns are global whereas
they are actually segments produced by local under-dense and over-dense regions, at least for 
relatively low mass disks. {\it These under-dense and over-dense regions act as perturbers,
keep alive the spiral activity and probe an entirely new regime in which the linear theory 
is no longer fully applicable.}

\item The dynamical response in the disk originates at the co-rotation resonance
between the perturbers and the surrounding stars. 
This implies that our features that wrap up and get
replaced statistically have angular speeds
that are not constant with location,
but decrease with radius. 
The feature speed slightly differs from the surrounding 
stars, explaining why the arms do not wind up while being at the same time long-lived. 

\item A modest overdensity corotating for at least 25 million years will excite a response 
in the disk that will lead
to a self-perpetuating pattern owing to non-linear effects.

\end{itemize}

We believe that there is already some
observational support for our picture.
In particular, our model predicts that there should exist red disk
galaxies with no gas but that may still have arms.  Galaxies such as this
have been discovered \citep{Mast10} and current
theories for spiral arm formation do not explain their morphologies.
In our model, the pattern speeds of the spiral features should
vary with location within a disk, as has been claimed for some
nearby spirals \citep{Foy10,Foy11,Meidt}.
We also predict that the number of arms in galaxies should depend on the
structural properties of the galaxy and that there will be a
correlation between the amplitude of the arms and how dynamically cold
the disk is, in the sense that colder disks should display more
prominent arms.  
These predictions are readily testable and
relevant observational
studies should provide strong tests of our model in
the forthcoming years.
 
\appendix

\section{Static potential for a thin exponential disk}
\noindent
To verify the importance
of self-gravity in the formation of arms,
we performed
an N-body simulation of a disk of stars treated as test particles. 
In order to treat the stars as test particles, we compute 
the forces acting on them
by introducing a static potential for a thin 
exponential disk of the form:
\begin{align}\label{ax}
a_x= & \, -\cos \theta \, \frac{\partial \Phi}{\partial R}   \notag \\
a_y= & \, -\sin \theta \, \frac{\partial \Phi}{\partial R}   \notag \\
a_z= & \, -\frac{\rm{G M_*}}{R_h^2}\exp{(-R/R_h)} \, \tanh  \left( \frac{z}{z_0} \right )
\end{align}
\noindent
where $\theta$ is the angle phase and
\begin{align}
\label{dphidr}
\frac{\partial \Phi}{\partial r}= & \, -\frac{GM_* z_0}{R_h^2} \ \exp{(-R/R_h)} 
\Big(\frac{1}{R_h}\Big) \ \rm{ln} \ \rm{cosh}\Big(\frac{z}{z_0}\Big) \notag \\
& \, -\frac{G M_*}{2 R_h} [I_0(y)K_1(y)-I_1(y)K_0(y)] \notag \\
& \, -\frac{G M_* y}{2 R_h^2}[I_0'(y)K_1(y)+I_0(y)K_1'(y)+ \notag \\
& \,  \ \ \ \ \ \ \ \ \ \ \ \ \ -I_1'(y)K_0(y)-I_1(y)K_0'(y)]
\end{align}

\noindent
The interactions between the perturbers and stars are computed 
directly, meaning that the stellar orbits are influenced by
these gravitational interactions, as in the simulations with
disk self-gravity presented above.

\section{Amplification parameter for a general disk galaxy}
\noindent
To interpret the results shown in \S 4.4 we consider the Toomre $X$-parameter
for a galaxy like that in our simulations, using
\begin{equation}\label{Xfactor}
X= \frac{\kappa^2}{2 \pi G \Sigma} \frac{R}{m}.
\end{equation}  
\noindent
In general,
the epicyclic frequency depends on 
the total angular frequency which is the sum of the angular frequencies of 
all the components of a galaxy, in particular the bulge, disk and dark halo,
according to
\begin{equation}
\Omega^2= \Omega_D^2+\Omega_B^2+\Omega_H^2 \, .
\end{equation}
\noindent
For an exponential disk:
\begin{equation}
\Omega_D^2=\frac{G M_D}{2 R_h^3} \Big[ I_0(y) K_0(y) -I_1(y) K_1(y) \Big]
\end{equation}
\noindent
where ${y=R/2R_h}$,
$R_h$ is the disk scale length and $M_D$ the disk total mass.
For bulge and halos described by Hernquist models,
the angular frequency of the bulge is
\begin{equation}
\Omega_B^2= \frac{G M_B}{(R+a_b)^2},
\end{equation}
\noindent 
where $a_b$ and $M_B$ are 
the bulge scale length and mass, respectively and for the halo
\begin{equation}
\Omega_H^2= \frac{G M_H}{(R+a_h)^2}
\end{equation}
where $a_h$ and $M_H$ are
the halo scale length and the mass, respectively.
\noindent
The amplification parameter  for a general disk galaxy becomes:
\begin{align}
X= & \, \frac{e^{2y}}{m} \Big( \Big[ \frac{M_B}{M_D} \frac{2y+3a_b/R_h}{(2y+a_b/R_h)^3}\Big] \notag \\
   & +  \,               \Big[ \frac{M_H}{M_D} \frac{2y+3a_h/R_h}{(2y+a_h/R_h)^3}\Big] \notag \\
   & +  \, \frac{y^2}{2} \Big( 3 I_1 K_0 -3 I_0 K_1 + I_1 K_2 -I_2 K_1 \Big) \notag \\
   & + \, 4y (I_0 K_0 - I_1 K_1) \Big)
\label{eqX} 
\end{align}
\noindent
where $m$ characterizes the Fourier coefficient $A_m$ and indicates the number of arms.
Note that substituting the values for the structural parameters of the
galaxy model adopted here into eq. (\ref{eqX}) gives an estimate of
the number of arms as $\sim 7$, in accord with our simulations.

\acknowledgments 
\noindent
We are grateful to the referee Jerry Sellwood for constructive suggestions and to 
Alar Toomre for much generous and wise advice. 
We thank Mark Reid and Debora Sijacki for a careful reading of the manuscript.
Numerical simulations were performed on the Odyssey supercomputer
at Harvard University.

\bibliographystyle{apj}
\bibliography{spiral}

\end{document}